\documentstyle[12pt]{article}

\title{The Rest-Frame Instant Form of Dynamics and Dirac's Observables.}
\author{\large Luca Lusanna \\[3mm]
\em Sezione INFN di Firenze, \\
\em Largo E.Fermi 2, 50125 Firenze, Italy\\
\em E-mail; lusanna@fi.infn.it}

\begin{document}

\maketitle

Talk given at the Int.Worshop ``Physical Variables in Gauge Theories",
Dubna 1999.

\vskip 2cm

Since our understanding of both general relativity and the standard
model of elementary particles  either with or without supersymmetry is
based on singular Lagrangians, whose associated Hamiltonian formalism
requires Dirac-Bergmann theory of constraints\cite{dirac}, it is very
difficult to identify which are the physical degrees of freedom to be
used in the description and interpretation of the fundamental
interactions. This is the problem of the physical observables in gauge
theories and general relativity.

 While behind the gauge
freedom of gauge theories proper there are Lie groups acting on some
internal space so that the measurable quantities must be gauge
invariant, the gauge freedom of theories invariant under
diffeomorphism groups of the underlying spacetime (general relativity,
string theory and reparametrization invariant systems of relativistic
particles) concerns the arbitrariness for the observer in the choice
of the definition of ``what is space and/or time" (and relative times
in the case of particles), i.e. of the definitory properties either of
spacetime itself or of the measuring apparatuses. This is the
classical mathematical background on which our understanding of the
quantum field theory of electromagnetic, weak and strong interactions
in the modern BRS formulation is based. The same is true for our
attempts to build quantum gravity notwithstanding our actual
incapacity to reconcile the influence of gravitational physics on the
existence and formulation of spacetime concepts with the basic ideas
of quantum theory, which requires a given absolute background
spacetime.

In Minkowski spacetime, one uses the covariant approach based on the
BRS symmetry which, at least at the level of the algebra of
infinitesimal gauge transformations, allows a regularization and
renormalization of the relevant theories inside the framework of local
quantum field theory. However, problems like the understanding of
finite gauge transformations and of the associated moduli spaces, the
Gribov ambiguity dependence on the choice of the function space for
the fields and the gauge transformations, the confinement of quarks,
the definition of relativistic bound states and how to put them among
the asymptotic states, the nonlocality of charged states in quantum
electrodynamics, not to speak of the conceptual and practical problems
posed by gravity, suggest that we should revisit the foundations of
our theories.

When we  will succeed to reformulate classical physics in terms of the
physical degrees of freedom hidden behind manifest gauge and/or
general covariance, the quantization  of the resulting theories will
require to abandon local field theory at the non-perturbative level
and to understand how to regularize and renormalize the Coulomb and
radiation gauges of electrodynamics to start with.

I revisited the classical Hamiltonian formulation of theories
described by singular Lagrangians trying to choose the mathematical
frameworks which at each step looked more natural to clarify the
physical interpretational problems by means of the use of suitable
adapted coordinates. In particular, after many years of dominance of
the point of view privileging manifest Lorentz, gauge and general
covariance at the price of loosing control on the physical degrees of
freedom and on their deterministic evolution (felt as a not necessary
luxury only source of difficulties and complications), I went back to
the old concept of Dirac observables, namely of those gauge invariant
deterministic variables which describe a canonical basis of measurable
quantities for the electromagnetic, weak and strong interactions in
Minkowski spacetime. Instead, in general relativity, due to the
problem of the individuation of the points of spacetime, measurable
quantities have a more complex identification, which coincides with
Dirac's observables (in any case indispensable for the treatment of
the Cauchy problem) only in a completely fixed gauge (total breaking
of general covariance).

See Ref.\cite{india} for a complete review of what has been understood
till now in the Hamiltonian framework of constraint theory, where the
physical observables of gauge theories are identified with the gauge
invariant Dirac observables (this definition can be extended to
theories invariant under diffeomorphisms like general relativity, but
I will not speak about this topic in this short review).

In the search of physical observables there are two primary problems:

A) The choice of the function space in which our classical field
theory is defined. The existence of the Gribov ambiguity in gauge
theories with the associated `cone over cone' structure of
singularities and stratification of both the space of solutions of the
equations of motion and of the associated constraint manifold in phase
space depends on such a choice. If the Gribov ambiguity is only a
mathematical obstruction one must work in special weighted Sobolev
spaces where it is absent. If, instead, there is some physics hidden
into it, one works in ordinary Sobolev spaces, but one has to face the
above singularities and stratification, which can constitute an
obstruction already at the classical level for the construction of
global Dirac observables. When this type of singularities can be
avoided, the search of Dirac observables is done by solving the
multitemporal equations and by looking for special Shanmugadhasan
canonical trasformations\cite{sha} to find special Darboux bases
adapted to the constraints of the theory and containing as a subset a
canonical basis of Dirac observables\cite {india,lu1}.

In the finite dimensional case general theorems connected with the Lie
theory of function groups ensure the existence of local Shanmugadhasan
canonical transformations from the original canonical variables $q^i$,
$p_i$, in terms of which the first class constraints (assumed globally
defined) have the form $\phi_{\alpha}(q,p)\approx 0$, to canonical
bases $P_{\alpha}$, $Q_{\alpha}$, $P_A$, $Q_A$, such that the
equations $P_{\alpha}\approx 0$ locally define the same original
constraint manifold (the $P_{\alpha}$ are an Abelianization of the
first class constraints); the $Q_{\alpha}$ are the adapted Abelian
gauge variables describing the gauge orbits (they are a realization of
the times $\tau_{\alpha}$ of the multitemporal equations in terms of
variables $q^i$, $p_i$); the $Q_A$, $P_A$ are an adapted canonical
basis of Dirac observables. These canonical transformations are the
basis of the Hamiltonian definition of the Faddeev-Popov measure of
the path integral\cite{fp} and give a trivialization of the BRS
construction of observables (the BRS method works when the first class
constraints may be Abelianized\cite{hennea}).

Putting equal to zero the Abelianized gauge variables one defines a
local gauge of the model. If a system with constraints admits one (or
more) global Shanmugadhasan canonical transformations, one obtains one
(or more) privileged global gauges in which the physical Dirac
observables are globally defined and globally separated from the gauge
degrees of freedom [for systems with a compact configuration space Q
this is generically impossible]. These privileged gauges (when they
exist) can be called generalized Coulomb gauges.

When there is reparametrization invariance of the original action
$S=\int dt L$, the canonical Hamiltonian vanishes and both kinematics
and dynamics are contained in the first class constraints describing
the system: these can be interpreted as generalized Hamilton-Jacobi
equations, so that the Dirac observables turn out to be the Jacobi
data. When there is a kinematical symmetry group, like the Galileo or
Poincar\'e groups, an evolution may be reintroduced by using the
energy generator as Hamiltonian.

Inspired by Ref.\cite{dira} the canonical reduction to noncovariant
generalized Coulomb gauges, with the determination of the physical
Hamiltonian as a function of a canonical basis of Dirac's observables,
has been achieved for the following isolated systems (for them one
asks that the 10 conserved generators of the Poincar\'e algebra are
finite so to be able to use group theory):

1) Relativistic particle mechanics. Its importance stems from the fact
that quantum field theory has no particle interpretation: this is
forced on it by means of the asymptotic states of the reduction
formalism which correspond to the quantization of independent one-body
systems described by relativistic mechanics. After a study of the
one-body problems corresponding to the basic wave equations, the
two-body DrozVincent-Todorov-Komar model [see Ref.\cite{india} for the
related bibliography] with an arbitrary action-at-a-distance
interaction instantaneous in the rest frame has been completely
understood both at the classical and quantum level \cite{longhi}. Its
study led to the identification of a class of canonical
transformations (utilizing the standard Wigner boost for timelike
Poincar\'e orbits) which allowed to understand how to define suitable
center-of-mass and relative variables (in particular a suitable
relative energy is determined by a combination of the two first class
constraints, so that the relative time variable is a gauge variable),
how to find a quasi-Shanmugadhasan canonical transformation adapted to
the constraint determining the relative energy, how to separate the
four, topologically disjoined, branches of the mass spectrum (it is
determined by the other independent combination of the constraints;
therefore, there is a distinct Shanmugadhasan canonical transformation
for each branch). At the quantum level it was possible to find four
physical scalar products, compatible with both the resulting coupled
wave equations (i.e. independent from the relative and the absolute
rest-frame times): they have been found as generalization of the two
existing scalar products of the Klein-Gordon equation: all of them are
non-local even in the limiting free case and differ among themselves
for the sign of the norm of states on different mass-branches. This
example shows that the physical scalar product knows the functional
form of the constraints. The wave functions of the quantized model can
be obtained from the solutions of a Bethe-Salpeter equation with the
same instantaneous potential  by multiplication for a delta function
containing the relative energy to exclude its spurious solutions (non
physical excitations in the relative energy). The extension of the
model to two pseudoclassical electrons and to an electron and a scalar
has been done and the first was used to get good fits to meson
spectra.

The previous canonical transformations were then extended to N free
particles described by N mass-shell first class constraints
$p^2_i-m^2_i\approx 0$ : N-1 suitable relative energies are determined
by N-1 combinations of the constraints (so that the conjugate N-1
relative times are gauge variables), while the remaining combination
determines the $2^N$ branches of the mass spectrum. The N gauge
freedoms associated with these N combinations of the first class
constraints are the freedom of the observer: i) in the choice of the
time parameter to be used for the overall evolution of the isolated
system; ii) in the choice of the description of the relative motions
with any given delay among the pairs of particles.

2) Both the open and closed Nambu string\cite{colomo}.

3) Yang-Mills theory with Grassmann-valued fermion fields \cite{lusa}
in the case of a trivial principal bundle over a fixed-$x^o$ $R^3$
slice of Minkowski spacetime with suitable Hamiltonian-oriented
boundary conditions; this excludes monopole solutions (to have them,
even if they have been not yet found experimentally, one needs a
nontrivial bundle and a variational principle formulated on the
bundle, because the gauge potentials on Minkowski spacetime are not
globally defined) and, since $R^3$ is not compactified, one has only
winding number and no instanton number. After a discussion of the
Hamiltonian formulation of Yang-Mills theory, of its group of gauge
transformations and of the Gribov ambiguity, the theory has been
studied in suitable  weighted Sobolev spaces where the Gribov
ambiguity is absent \cite{vince} and the global color charges are well
defined. The global Dirac observables are the transverse quantities
${\vec A}_{a\perp} (\vec x,x^o)$, ${\vec E}_{a\perp}(\vec x,x^o)$ and
fermion fields dressed with Yang-Mills (gluonic) clouds. The nonlocal
and nonpolynomial (due to the presence of classical Wilson lines along
flat geodesics) physical Hamiltonian has been obtained: it is nonlocal
but without any kind of singularities, it has the correct Abelian
limit if the structure constants are turned off, and it contains the
explicit realization of the abstract Mitter-Viallet metric.

4) The Abelian and non-Abelian SU(2)
Higgs models with fermion fields\cite{lv}, where the
symplectic decoupling is a refinement of the concept of unitary gauge.
There is an ambiguity in the solutions of the Gauss law constraints, which
reflects the existence of disjoint sectors of solutions of the Euler-Lagrange
equations of Higgs models. The physical Hamiltonian and Lagrangian of  the
Higgs phase have been found; the self-energy turns out to be local and
contains a local four-fermion interaction.

5) The standard SU(3)xSU(2)xU(1) model of elementary particles\cite{lv3}
with \hfill\break Grassmann-valued fermion fields.
The final reduced Hamiltonian contains nonlocal self-energies for the
electromagnetic and color interactions, but ``local ones" for the weak
interactions implying the nonperturbative emergence of 4-fermions interactions.

\vskip 2cm

B) All the physical systems defined in the flat Minkowski spacetime,
have the global Poincare' symmetry. This suggests to study the
structure of the constraint manifold  from the point of view of the
orbits of the Poincare' group. If $p^{\mu}$ is the total momentum of
the system, the constraint manifold has to be divided in four strata
(some of them may be absent for certain systems) according to whether
$p^2 > 0$, $p^2=0$, $p^2 < 0$ or $p^{\mu}=0$. Due to the different
little groups of the various Poincare' orbits, the gauge orbits of
different sectors will not be diffeomorphic. Therefore the constraint
manifold is a stratified manifold and the gauge foliations of
relativistic systems are nearly never nice, but rather one has to do
with singular foliations. For an acceptable relativistic system the
stratum $p^2 < 0$ has to be absent to avoid tachyons. To study the
strata $p^2=0$ and $p^{\mu}=0$ one has to add these relations as extra
constraints.

For all the strata the next step is to do a canonical transformation
from the original variables to a new set consisting of center-of-mass
variables $x^{\mu}$, $p^{\mu}$ and of variables relative to the center
of mass. Let us now consider the stratum $p^2 > 0$. By using the
standard Wigner boost $L^{\mu}_{\nu}(p, {\buildrel \circ \over p})$
($p^{\mu}=L^{\mu}_{\nu}(p,{\buildrel \circ \over p}){\buildrel \circ
\over p}^{\nu}$, ${\buildrel \circ \over p}^{\mu}=\eta
\sqrt {p^2} (1;\vec 0 )$, $\eta = sign\, p^o$), one boosts the relative
variables at rest.
The new variables are still canonical and the base is completed by $p^{\mu}$
and by a new center-of-mass coordinate ${\tilde x}^{\mu}$, differing from
$x^{\mu}$ for spin terms. The variable ${\tilde x}^{\mu}$ has complicated
covariance properties; instead the new relative variables are either Poincare'
scalars or Wigner spin-1 vectors, transforming under the group O(3)(p) of
the Wigner rotations induced by the Lorentz transformations. A final
canonical transformation\cite{longhi}, leaving fixed the relative variables,
sends the center-of-mass coordinates ${\tilde x}^{\mu}$, $p^{\mu}$ in the new
set $p\cdot {\tilde x}/\eta \sqrt {p^2}=p\cdot x/\eta \sqrt {p^2}$ (the time
in the rest frame), $\eta \sqrt {p^2}$ (the total mass), $\vec k =\vec p
/\eta \sqrt {p^2}$ (the spatial components of the 4-velocity $k^{\mu}=
p^{\mu}/\eta \sqrt {p^2}$, $k^2=1$), $\vec z=\eta \sqrt {p^2}( {\vec
{\tilde x}}-{\tilde x}^o\vec p/p^o)$. $\vec z$ is a noncovariant
center-of-mass canonical 3-coordinate  multiplied by the total mass:
it is the classical analog of the Newton-Wigner position operator
(like it, $\vec z$ is covariant only under the little group O(3)(p) of
the timelike Poincar\'e orbits). Analoguous considerations could be
done for the other sectors.

The nature of the relative variables depends on the system. The first class
constraints, once rewritten in terms of the new variables, can be manipulated
to find suitable global and Lorentz scalar Abelianizations.
Usually there is a combination of the constraints which determines $\eta
\sqrt {p^2}$, i.e. the mass spectrum, so that the time in the rest frame
$p\cdot x/\eta \sqrt {p^2}$ is the conjugated Lorentz scalar gauge variable.
The other constraints eliminate some of the relative variables (in particular
the relative energies for systems of interacting relativistic particles and the
string): their conjugated coordinates (the relative times) are the other gauge
variables: they are identified with a possible set of time parameters by the
multitemporal equations. The Dirac observables (apart from the center-of-mass
ones $\vec k$ and $\vec z$) have to be extracted from the
remaining relative variables and the construction shows that they will be
either Poincare' scalars or Wigner covariant objects.
In this way in each stratum preferred global Shanmugadhasan canonical
transformations are identified, when no other kind of obstruction to
globality is present inside the various strata.

To covariantize the description of the previous reduced systems, which
is valid in Minkowski spacetime with Cartesian coordinates, again the
starting point was given by Dirac\cite{dirac} with his reformulation
of classical field theory on spacelike hypersurfaces foliating
Minkowski spacetime $M^4$ [the foliation is defined by an embedding
$R\times
\Sigma \rightarrow M^4$, $(\tau ,\vec
\sigma ) \mapsto z^{(\mu )}(\tau ,\vec \sigma )\in \Sigma_{\tau}$, with
$\Sigma$ an abstract 3-surface diffeomorphic to $R^3$, with
$\Sigma_{\tau}$ its copy embedded in $M^4$ labelled by the value
$\tau$ (the Minkowski flat indices are $(\mu )$; the scalar ``time"
parameter $\tau$ labels the leaves of the foliation, $\vec \sigma$ are
curvilinear coordinates on $\Sigma_{\tau}$ and $(\tau ,\vec \sigma )$
are $\Sigma_{\tau}$-adapted holonomic coordinates for $M^4$); this is
the classical basis of Tomonaga-Schwinger quantum field theory]. In
this way one gets a parametrized field theory with a covariant 3+1
splitting of Minkowski spacetime and already in a form suited to the
transition to general relativity in its ADM canonical formulation. The
price is that one has to add as new independent configuration
variables  the embedding coordinates $z^{(\mu )}(\tau ,\vec \sigma )$
of the points of the spacelike hypersurface $\Sigma_{\tau}$ [the only
ones carrying Lorentz indices] and then to define the fields on
$\Sigma_{\tau}$ so that they know the hypersurface $\Sigma_{\tau}$ of
$\tau$-simultaneity [for a Klein-Gordon field $\phi (x)$, this new
field is $\tilde \phi (\tau ,\vec \sigma )=\phi (z(\tau ,\vec \sigma
))$: it contains the nonlocal information about the embedding]. Then
one rewrites the Lagrangian of the given isolated system in the form
required by the coupling to an external gravitational field, makes the
previous 3+1 splitting of Minkowski spacetime and interpretes all the
fields of the system as the new fields on $\Sigma_{\tau}$ (they are
Lorentz scalars, having only surface indices). Instead of considering
the 4-metric as describing a gravitational field (and therefore as an
independent field as it is done in metric gravity, where one adds the
Hilbert action to the action for the matter fields), here one replaces
the 4-metric with the the induced metric $g_{ AB}[z]
=z^{(\mu )}_{A}\eta_{(\mu )(\nu )}z^{(\nu )}_{B}$ on
$\Sigma_{\tau}$ [a functional of $z^{(\mu )}$;
here we use the notation $\sigma^{A}=(\tau ,\sigma^{r})$; $z^{(\mu )}_{A}=
\partial z^{(\mu )}/\partial \sigma^{A}$ are flat tetrad fields on Minkowski
spacetime with the $z^{(\mu )}_r$'s tangent to $\Sigma_{\tau}$]
and considers the embedding coordinates $z^{(\mu )}(\tau ,\vec \sigma )$ as
independent fields [this is not possible in metric gravity, because in curved
spacetimes $z^{\mu}_{A}\not= \partial z^{\mu}/\partial \sigma^{A}$ are not
tetrad fields so that holonomic coordinates $z^{\mu}(\tau ,\vec \sigma )$
do not exist]. From this Lagrangian,
besides a Lorentz-scalar form of the constraints of the given system,
we get four extra primary first class constraints\hfill\break

${\cal H}
_{(\mu )}(\tau ,\vec \sigma )=\rho_{(\mu )}(\tau ,\vec \sigma )-l_{(\mu )}(\tau
,\vec \sigma )T_{sys}^{\tau\tau}(\tau ,\vec \sigma )-z_{r
(\mu )}(\tau ,\vec \sigma )T_{sys}^{\tau r}(\tau ,\vec \sigma )
\approx 0$\hfill\break

[here $T_{sys}^{\tau\tau}(\tau ,\vec \sigma )$, $T_{sys}
^{\tau r}(\tau ,\vec \sigma )$, are the components of
the energy-momentum tensor in the holonomic coordinate system,
corresponding to the energy- and momentum-density of the
isolated system; one has $\lbrace {\cal H}_{(\mu )}(\tau ,\vec \sigma ),
{\cal H}_{(\nu )}(\tau ,{\vec \sigma}^{'}) \rbrace =0$]
implying the independence of the description from the choice of the 3+1
splitting, i.e. from the choice of the foliation with spacelike hypersufaces.

In special relativity, it is convenient to restrict ourselves to
arbitrary spacelike hyperplanes $z^{(\mu )}(\tau ,\vec \sigma
)=x^{(\mu )}_s(\tau )+ b^{(\mu )}_{r}(\tau ) \sigma^{r}$. Since they
are described by only 10 variables, after this restriction we remain
only with 10 first class constraints determining the 10 variables
conjugate to the hyperplane in terms of the variables of the system:
\hfill\break

${\cal H}^{(\mu )}(\tau )=p^{(\mu )}_s-p^{(\mu )}
_{(sys)}\approx 0$, ${\cal H}^{(\mu )(\nu )}(\tau )=S^{(\mu )(\nu )}
_s-S^{(\mu )(\nu )}_{(sys)}\approx 0$. \hfill\break

The 20 variables for the phase space description of a hyperplane are:
\hfill\break
i) $x^{(\mu )}_s(\tau ), p^{(\mu )}_s$, parametrizing the origin of
the coordinates on the family
of spacelike hyperplanes. The four constraints ${\cal H}^{(\mu )}(\tau )
\approx 0$ say that $p_s^{(\mu )}$ is determined by the 4-momentum of the
isolated system.\hfill\break
ii) $b^{(\mu )}_A(\tau )$ (with the $b^{(\mu )}_r(\tau )$'s being three
orthogonal spacelike unit vectors generating the fixed $\tau$-independent
timelike unit normal $b^{(\mu )}_{\tau}=l^{(\mu )}$ to the hyperplanes)
and $S^{(\mu )(\nu )}_s=-S^{(\nu )(\mu )}_s$ with the orthonormality
constraints $b^{(\mu )}_A\, {}^4\eta_{(\mu )(\nu )} b^{(\nu )}_B={}^4\eta_{AB}$
[enforced by assuming the Dirac brackets
$\{ S^{(\mu )(\nu )}_s,b^{(\rho )}_A \}={}^4\eta^{(\rho
)(\nu )} b^{(\mu )}_A-{}^4\eta^{(\rho )(\mu )} b^{(\nu )}_A$,
$\{ S^{(\mu )(\nu )}_s,S^{(\alpha )(\beta )}_s \} =C^{(\mu )(\nu )(\alpha
)(\beta )}_{(\gamma )(\delta )} S^{(\gamma )(\delta )}_s$
with $C^{(\mu )(\nu
)(\alpha )(\beta )}_{(\gamma )(\delta )}$ the structure constants of the
Lorentz algebra]. In these variables there are hidden six independent
pairs of degrees of freedom. The six constraints ${\cal H}^{(\mu )(\nu )}
(\tau )\approx 0$ say that
$S_s^{(\mu )(\nu )}$ coincides the spin tensor of the isolated system.
Then one has that $p^{(\mu )}_s$, $J^{(\mu )(\nu )}_s=x
^{(\mu )}_sp^{(\nu )}_s-x^{(\nu )}_sp^{(\mu )}_s+S^{(\mu )(\nu )}_s$, satisfy
the algebra of the Poincar\'e group.

Let us remark that,
for each configuration of an isolated system there is a privileged
family of hyperplanes (the Wigner hyperplanes orthogonal to $p^{(\mu )}_s$,
existing when $ p^2_s > 0$) corresponding to the intrinsic rest-frame
of the isolated system. If we choose these hyperplanes with suitable
gauge fixings, we remain with only  the four constraints ${\cal H}^{(\mu )}(\tau
)\approx 0$, which can be rewritten as\hfill\break
\hfill\break
$\sqrt{p^2_s} \approx [invariant\, mass\, of\, the\,
isolated\, system\, under\, investigation]= M_{sys}$; \hfill\break
${\vec p}_{sys}=[3-momentum\, of\, the\, isolated\, system\,
inside\, the\, Wigner\, hyperplane]\approx 0$.\hfill\break
\hfill\break
There is no more a restriction on $p_s^{(\mu )}$, because
$u^{(\mu )}_s(p_s)=p^{(\mu )}_s/p^2_s$ gives the orientation of the Wigner
hyperplanes containing the isolated system with respect to an arbitrary
given external observer.

In this special gauge we have $b^{(\mu )}_A\equiv L^{(\mu )}{}_A(p_s,{\buildrel
\circ \over p}_s)$ (the standard Wigner boost for timelike Poincar\'e orbits),
$S_s^{(\mu )(\nu )}\equiv S_{system}^{(\mu )(\nu )}$, and the only
remaining canonical variables are the noncovariant Newton-Wigner-like
canonical ``external" center-of-mass
coordinate ${\tilde x}^{(\mu )}_s(\tau )$ (living on the
Wigner hyperplanes) and $p^{(\mu )}_s$.
Now 3 degrees of freedom of the isolated system [an ``internal"
center-of-mass 3-variable ${\vec \sigma}_{sys}$ defined inside the Wigner
hyperplane and conjugate to ${\vec p}_{sys}$] become gauge variables [the
natural gauge fixing is ${\vec \sigma}_{sys}\approx 0$, so that it coincides
with the origin $x^{(\mu )}_s(\tau )=z^{(\mu )}(\tau ,\vec \sigma =0)$ of the
Wigner hyperplane], while the ${\tilde x}^{(\mu )}$
is playing the role of a kinematical external
center of mass for the isolated system and may be interpreted as a decoupled
observer with his parametrized clock (point particle clock).
All the fields living on the Wigner hyperplane are now either Lorentz scalar
or with their 3-indices transformaing under Wigner rotations (induced by Lorentz
transformations in Minkowski spacetime) as any Wigner spin 1 index.

One obtains
in this way a new kind of instant form of the dynamics (see Ref.\cite{dira2}),
the  ``Wigner-covariant 1-time rest-frame instant form"\cite{lus1} with a
universal breaking of Lorentz covariance.
It is the special relativistic generalization of
the nonrelativistic separation of the center of mass from the relative motion
[$H={{ {\vec P}^2}\over {2M}}+H_{rel}$]. The role of the center of mass is
taken by the Wigner hyperplane, identified by the point ${\tilde x}^{(\mu )}
(\tau )$ and by its normal $p^{(\mu )}_s$. The
invariant mass $M_{sys}$ of the system replaces the nonrelativistic  Hamiltonian
$H_{rel}$ for the relative degrees of freedom, after the addition of the
gauge-fixing $T_s-\tau \approx 0$ [identifying the time parameter $\tau$,
labelling the leaves of the foliation,  with
the Lorentz scalar time of the center of mass in the rest frame,
$T_s=p_s\cdot {\tilde x}_s/M_{sys}$; $M_{sys}$  generates the
evolution in this time].

The determination of ${\vec \sigma}_{sys}$ may be done with the group
theoretical methods of Ref.\cite{pauri}: given  the ``internal"
realization on the phase space of a given system of the ten Poincar\'e
generators [it is determined by the previous constraints] one can
build three 3-position variables only in terms of them, which in our
case of a system on the Wigner hyperplane with ${\vec p}_{sys}\approx
0$ are: i) a canonical center of mass (the ``internal" center of mass
${\vec
\sigma}_{sys}$); ii) a noncanonical M\o ller center of energy ${\vec
\sigma}^{(E)}_{sys}$; iii) a noncanonical Fokker-Pryce center of
inertia ${\vec
\sigma}^{(FP)}_{sys}$. Due to ${\vec p}_{sys}\approx 0$, we have
${\vec \sigma}_{sys} \approx {\vec
\sigma}^{(E)}_{sys}
\approx {\vec \sigma}^{(FP)}_{sys}$. By adding the gauge fixings
${\vec \sigma}_{sys}\approx 0$ one can show that the origin $x_s^{(\mu
)}(\tau )$ becomes  simultaneously the Dixon center of mass of an
extended object and both the Pirani and Tulczyjew centroids (see Ref.
\cite{mate} for the application of these methods to find the center of
mass of a configuration of the Klein-Gordon field after the
preliminary work of Ref.\cite{lon}). With similar methods one can
construct three ``external" collective positions (all located on the
Wigner hyperplane) from the rest-frame instant form realization of the
``external" Poincar\'e group: i) the ``external" canonical
noncovariant center of mass ${\tilde x}_s^{(\mu )}$; ii) the
``external" noncanonical and noncovariant M\o ller center of energy
$R^{(\mu )}_s$; iii) the ``external" covariant noncanonical
Fokker-Pryce center of inertia $Y^{(\mu )}_s$ (when there are the
gauge fixings ${\vec
\sigma}_{sys}\approx 0$ it also coincides with the origin $x^{(\mu
)}_s$). It turns out that the Wigner hyperplane is the natural setting
for the study of the Dixon multipoles of extended relativistic
systems\cite{dixon} and for defining the canonical relative variables
with respect to the center of mass. The Wigner hyperplane with its
natural Euclidean metric structure offers a natural solution to the
problem of boost for lattice gauge theories and realizes explicitly
the machian aspect of dynamics that only relative motions are
relevant.

The isolated systems till now analyzed to get their rest-frame
Wigner-covariant generalized
Coulomb gauges [i.e. the subset of global Shanmugadhasan canonical bases,
which, for each Poincar\'e stratum, are also adapted to the geometry of the
corresponding Poincar\'e orbits with their little groups; these special bases
can be named Poincar\'e-Shanmugadhasan bases for the given Poincar\'e stratum
of the presymplectic constraint manifold (every stratum requires an independent
canonical reduction); till now only the main stratum with
$p^2$ timelike and $W^2\not= 0$ has been investigated] are:

a) The system of N scalar particles with Grassmann electric charges
plus the electromagnetic field \cite{lus1}. The starting configuration
variables are a 3-vector ${\vec \eta}_i(\tau )$ for each particle [$x^{(\mu )}
_i(\tau )=z^{(\mu )}(\tau ,{\vec \eta}
_i(\tau ))$] and the electromagnetic gauge potentials
$A_{A}(\tau ,\vec \sigma )={{\partial z^{(\mu )}(\tau ,\vec \sigma )}
\over {\partial \sigma^{A}}} A_{(\mu )}(z(\tau ,\vec \sigma ))$,
which know  the embedding of
$\Sigma_{\tau}$ into $M^4$. One has to choose the sign of the energy of each
particle, because there are not mass-shell constraints (like $p_i^2-m^2_i\approx
0$) among the constraints of this formulation, due to the fact that one has only
three degrees of freedom for particle, determining the intersection of a
timelike trajectory and of the spacelike hypersurface $\Sigma_{\tau}$. For
each choice of the sign of the energy of the N particles, one describes only one
of the $2^N$ branches of the mass spectrum of the manifestly covariant approach
based on the coordinates $x^{(\mu )}_i(\tau )$, $p^{(\mu )}_i(\tau )$,
i=1,..,N, and on
the constraints $p^2_i-m^2_i\approx 0$ (in the free case). In this way, one
gets a description of relativistic particles with a given sign of the energy
with consistent couplings to fields and valid independently from the quantum
effect of pair production [in the manifestly covariant approach, containing
all possible branches of the particle mass spectrum, the classical counterpart
of pair production is the intersection of different branches deformed by the
presence of interactions]. The final Dirac's observables are: i) the transverse
radiation field variables ${\vec A}_{\perp}$, ${\vec E}_{\perp}$;
ii) the particle canonical variables ${\vec \eta}_i(\tau )$, ${{\vec \kappa}}
_i(\tau )$, dressed with a Coulomb cloud. The physical Hamiltonian contains the
mutual instantaneous Coulomb potentials extracted from field theory
and there is a regularization of the Coulomb self-energies due to the
Grassmann character of the electric charges $Q_i$ [$Q^2_i=0$]. In
Ref.\cite{lus2} there is the study of the Lienard-Wiechert potentials
and of Abraham-Lorentz-Dirac equations in this rest-frame Coulomb
gauge and also scalar electrodynamics is reformulated in it. Also the
rest-frame 1-time relativistic statistical mechanics has been
developed \cite{lus1}. The extraction of the Darwin potential from the
Lienard-Wiechert solution is nearly accomplished \cite{horace}. A
general study of the relativistic center od mass, of the rotational
kinematics and of Dixon multipolar expansions\cite{dixon} is under
investigation\cite{iten}.

b) The system of N scalar particles with Grassmann-valued color charges plus
the color SU(3) Yang-Mills field\cite{lus3}:
it gives the pseudoclassical description of the
relativistic scalar-quark model, deduced from the classical QCD Lagrangian and
with the color field present. The physical invariant mass of the system is
given in terms of the Dirac observables. From the reduced Hamilton equations
the second order equations of motion both for the reduced transverse color
field and the particles are extracted. Then, one studies  the N=2
(meson) case. A special form of the requirement of having only color singlets,
suited for a field-independent quark model, produces a ``pseudoclassical
asymptotic freedom" and a regularization of the quark self-energy. With these
results one can covariantize the bosonic part of the standard model given in
Ref.\cite{lv3}.

c) The system of N spinning particles of definite energy [$({1\over
2},0)$ or $(0,{1\over 2})$ representation of SL(2,C)] with Grassmann
electric charges plus the electromagnetic field\cite{biga} and that of
a Grassmann-valued Dirac field plus the electromagnetic field (the
pseudoclassical basis of QED) \cite{bigaz}. In both cases there are
geometrical complications connected with the spacetime description of
the path of electric currents and not only of their spin structure,
suggesting a reinterpretation of the supersymmetric scalar multiplet
as a spin fibration with the Dirac field in the fiber and the
Klein-Gordon field in the base; a new canonical decomposition of the
Klein-Gordon field into center-of-mass and relative variables
\cite{lon,mate} will be helpful to clarify these problems. After their
solution and after having obtained the description of Grassmann-valued
chiral fields the rest-frame form of the full standard $SU(3)\times
SU(2)\times U(1)$ model can be achieved.

d) The study of the definition of collective and relative variables
for the Klein-Gordon field, initiated in Refs.\cite{lon}, has been
reformulated in the rest-frame instant form\cite{mate} with a
discussion of how to find the canonical internal center of mass of the
field configuration. Also Dixon's multipolar expansions are studied at
the Hamiltonian level on the Wigner hyperplanes. Now the same
problematic is under investigation for the electromagnetic
field\cite{ass}.

e) The relativistic perfect fluids\cite{milla}.

As shown in Refs.\cite{lus1,lusa}, the rest-frame instant form of
dynamics automatically gives a physical ultraviolet cutoff in the
spirit of Dirac and Yukawa: it is the M$\o$ller radius\cite{mol} $\rho
=\sqrt{-W^2}/p^2=|\vec S|/\sqrt{p^2}$ ($W^2=-p^2{\vec S}^2$ is the
Pauli-Lubanski Casimir when $p^2 > 0$), namely the classical intrinsic
radius of the worldtube, around the covariant noncanonical
Fokker-Pryce center of inertia $Y^{(\mu )}$, inside which the
noncovariance of the canonical center of mass ${\tilde x}^{\mu}$ is
concentrated. At the quantum level $\rho$ becomes the Compton
wavelength of the isolated system multiplied its spin eigenvalue
$\sqrt{s(s+1)}$ , $\rho \mapsto \hat \rho = \sqrt{s(s+1)} \hbar
/M=\sqrt{s(s+1)} \lambda_M$ with $M=\sqrt{p^2}$ the invariant mass and
$\lambda_M=\hbar /M$ its Compton wavelength. Therefore, the criticism
to classical relativistic physics, based on quantum pair production,
concerns the testing of distances where, due to the Lorentz signature
of spacetime, one has intrinsic classical covariance problems: it is
impossible to localize the canonical center of mass ${\tilde x}^{\mu}$
adapted to the first class constraints of the system (also named Pryce
center of mass and having the same covariance of the Newton-Wigner
position operator) in a frame independent way. For more details see
Ref.\cite{india}.

In conclusion, the best set of canonical coordinates adapted to the constraints
and to the geometry of Poincar\'e orbits in Minkowski spacetime
and naturally predisposed to the
coupling to canonical tetrad gravity is emerging for the electromagnetic, weak
and strong interactions with matter described either by fermion fields or by
relativistic particles with a definite sign of the energy.

After having studied the canonical reduction in the 3-orthogonal gauge
of a new formulation of tetrad gravity\cite{india,russo}  and its
deparametrization to the rest-frame instant form of dynamics in
Minkowski spacetime when the Newton constant is switched off, the main
tasks for the future are:

A) Make the canonical quantization of scalar electrodynamics in the
rest-frame instant form on the Wigner hyperplanes, which should lead
to a particular realization of Tomonaga-Schwinger quantum field
theory, avoiding the no-go theorems of Refs.\cite{torre}. The main
problems to be investigated are\hfill\break
 $\quad$ i) the use of the
wave functions of the quantization of positive energy scalar
particles\cite{lam} to define Tomonaga-Schwinger asymptotic states and
a LSZ reduction formalism [since Fock states do not constitute a
Cauchy problem for the field equations, because an in (or out)
particle can be in the absolute future of another one due to the
tensor product nature of these asymptotic states, bound state
equations like the Bethe-Salpeter one have spurious solutions which
are excitations in relative energies, the variables conjugate to
relative times]. Moreover, it should be possible to include bound
states among the asymptotic states.\hfill\break
 $\quad$ ii) The search
of a Schroedinger-like equation for bound states by using the
Schwinger-Dyson equations but avoiding the Bethe-Salpeter equation
with its spurious solutions \cite{india} [see Refs.\cite{schw, prot}
for the nonrelativistic case).\hfill\break
 $\quad$ iii) Use the M\o
ller radius as a physical ultraviolet cutoff for the point splitting
technique.\hfill\break
 $\quad$ iv) See how to use the results of
Refs.\cite{lavelle} about the infrared dressing of asymptotic states
in S matrix theory to avoid the `infraparticle' problem\cite{buc}.

B) Study the linearization of tetrad gravity in the 3-orthogonal gauge
to reformulate the theory of gravitational waves in this gauge.

C) Study the N-body problem in tetrad gravity at the lowest order in
the Newton constant (action-at-a-distance plus linearized tetrad
gravity).

D) Begin the study of the standard model of elementary particles
coupled to tetrad gravity starting from the Einstein-Maxwell system.

\end{document}